# Antiferromagnetic Domain Wall Engineering in Chromium Films

J. M. Logan[1,a)], H. C. Kim[1], D. Rosenmann[2], Z. Cai[3], R. Divan[2], O. G. Shpyrko[4] and E. D. Isaacs[1,5]

[1] *The University of Chicago Department of Physics, Chicago, Illinois 60637, USA*

[2] *Center for Nanoscale Materials, Argonne National Laboratory, Argonne, Illinois 60439, USA*

[3] *Advanced Photon Source, Argonne National Laboratory, Argonne, Illinois 60439, USA*

[4] *The University of California San Diego, La Jolla, California 92093, USA*

[5] *Argonne National Laboratory, Argonne, Illinois 60439, USA*

We have engineered an antiferromagnetic domain wall by utilizing a magnetic frustration effect of a thin iron cap layer deposited on a chromium film. Through lithography and wet etching we selectively remove areas of the Fe cap layer to form a patterned ferromagnetic mask over the Cr film. Removing the Fe locally removes magnetic frustration in user-defined regions of the Cr film. We present x-ray microdiffraction microscopy results confirming the formation of a 90° spin-density wave propagation domain wall in Cr. This domain wall nucleates at the boundary defined by our Fe mask.

Antiferromagnets play a critical and growing role in much of contemporary condensed matter physics and technology. Although traditionally used in magnetic storage devices (e.g., readheads in hard drives) for their supporting role as pinning layers, antiferromagnetic materials are now being used in emerging technologies as the principal layer governing transport properties of the device, such as spin valves utilizing antiferromagnetic tunneling anisotropic magnetoresistance [1]. With the emergence of these new technologies, the ability to understand

---

a) Electronic Mail: jmlogan@uchicago.edu

and control antiferromagnetic domains and domain walls is becoming increasingly important. How these domain walls form, move, and affect electron transport is at the heart of understanding intrinsic properties of antiferromagnetic materials. Currently, much less is known about antiferromagnetic domain walls than their ferromagnetic counterparts, but recent studies on the elemental antiferromagnet chromium are revealing new and potentially useful properties of antiferromagnetic domains and domain walls that differ dramatically from those of ferromagnets [2-3]. For example, Jaramillo et al., found that the electrical interface resistance of an antiferromagnetic domain wall in Cr is nearly two orders of magnitude larger than that typical of ferromagnets [3].

Obtaining local information about antiferromagnetic domains requires sophisticated techniques, owing to their vanishing net magnetic moment [4]. These domains form in irregular and widely varying shapes and sizes, further confounding attempts at understanding their local properties. Additionally, antiferromagnetic domains are weakly pinned and subject to thermal or even quantum mechanical fluctuations [5-6]. It has been recognized that if antiferromagnetism is to play a larger role in electronic and magnetic devices, this limitation must be overcome by engineering more temporally stable domain walls [6]. The ability to introduce artificial pinning centers for domain wall nucleation in antiferromagnets allows new possibilities for understanding and ultimate control of these systems.

The itinerant spin-density-wave (SDW) antiferromagnetism in chromium has been studied extensively in both bulk and film samples [7-8]. Below its Néel temperature of 311K, bulk Cr exhibits an incommensurate SDW characterized by a spin polarization vector $\boldsymbol{S}$ and propagation wavevector $\boldsymbol{Q}$ whose magnitude is given by $Q = 2\pi/a(1\pm\delta)$, where $a$ is the lattice constant and $\delta$ the incommensurability parameter. In strain-free samples, Cr forms roughly equal populations of SDW domains that propagate along any one of the three equivalent b.c.c (100) crystallographic axes. Accompanying the SDW is a charge-density wave (CDW), a combination of ionic and itinerant charge modulation, propagating along the same ***Q*** direction but with half the wavelength. Due to the greater x-ray scattering intensity and non-dependence on spin polarization of the CDW, our investigations of the antiferromagnetic SDW Q domains consisted of probing the CDW superstructure concomitant with the SDW.

Our technique relies on two previous findings for (001) oriented Cr films. The first is that Cr films of thicknesses ranging from 50 to 300 nm exhibit an orientational pinning such that a single Q state exists with ***Q*** perpendicular to the film surface [9]. The second is that films thicker than 200 nm with a ferromagnetic cap layer undergo a complete ***Q*** vector reorientation such that for every domain ***Q*** orients parallel to the film surface along one of the two in-plane cubic axes [10]. The prevailing theory is that the ***Q*** vector reorientation minimizes spin frustration arising from steps and interdiffusion at the interface between Cr and the ferromagnetic cap layer [10].

By combining these two effects we have devised a method for producing artificial Q domain walls. Our method for producing antiferromagnetic domain walls begins with deposition of a ferromagnetic layer of Fe on a Cr film to induce reorientation of ***Q*** in Cr. Next, a lithographically patterned photoresist mask is created over the Fe. Selective wet etching of the Fe layer is then used to create a patterned Fe mask thereby generating antiferromagnetic domain walls in the underlying Cr layer. We confirmed the presence of antiferromagnetic domain walls in Cr along the pattern boundaries defined by the Fe mask and measured their widths using scanning x-ray microdiffraction microscopy.

High quality, single crystal Cr films were grown epitaxially on (001) oriented MgO substrates through dc magnetron sputtering with growth conditions similar to those found in the literature [11,2]. Typical base pressures during deposition were 3 x $10^{-8}$ Torr. A 30 nm buffer layer of Cr was grown at 650°C, followed by 270 nm of Cr deposited at 200°C and then annealed in vacuum at 700°C. Next, an Fe cap layer of 8 nm thickness was deposited at 300°C, and finally a 15 nm layer of Au was deposited to protect the film from oxidation. All depositions were performed sequentially without exposing the sample to any gas other than argon. The mosaicity of our Cr film was characterized by the 0.28° FWHM rocking curve of the (0,0,2) Bragg peak. The Néel temperature, as determined from the electrical resistivity anomaly of Cr, was centered at 297 K. Effects of antiferromagnetic ordering on the resistivity did not fully disappear until 305 K.

Following the deposition, the Fe and Au layers were prepared for etching from a portion of the surface by covering the sample with a positive photoresist and patterning the mask with an optical laser-writer. Resist development was followed by wet chemical etching to remove only the Fe and Au layers from the patterned parts of the sample. This process left a part of the Cr sample capped with Fe and Au, while the uncapped remainder consisted solely of Cr. For simplicity, the first region will be referred to henceforth as the *capped* region and the other as the *uncapped* region. In the uncapped region, where the magnetic frustration was removed with the removal of the Fe layer, we expected a single domain configuration with *Q* normal to the film plane. In the capped region, where the magnetic frustration remained, we expected domains oriented with *Q* in one of the two possible in-plane directions. Figure 1 shows a schematic of the expected *Q* directions within our sample as well as a SEM image of the sample after patterning.

The experiments were performed at Sector 2ID-D of the Advanced Photon Source at Argonne National Laboratory. X-rays of energy 10.1 keV were selected by a Si(111) monochromator and were focused to a sub-500 nm beam size through a Fresnel zone plate, with the desired harmonics selected through an order-sorting aperture. The sample was mounted with GE varnish on the cold finger of a liquid helium flow cryostat and made accessible to x-rays with a beryllium dome.

Bragg scans along the (0,0,$L$) direction in reciprocal space were taken at a location 400 μm on either side of the capped-uncapped boundary with an unfocused (200 x 200) μm$^2$ beam. Figure 2(a) shows scans in the uncapped region which reveal a temperature dependent peak located initially at (0,0,1.901) at 4 K (top red curve) which becomes much weaker and shifts to (0,0,1.923) upon warming to RT (top blue curve). This peak corresponds to the (0,0,2-2$\delta$) CDW satellite which has an increased value of $\delta$ and scattered intensity with decreasing temperature [12]. The dashed black line is a Gaussian fit to the tail of the (0,0,2) Bragg reflection. The same Gaussian parameters were used to fit the data collected at 4 K and RT. The bottom red and blue curves show the data collected at 4 K and RT, respectively, after subtracting this Gaussian function to remove the tail of the (0,0,2) Bragg reflection. The temperature dependence of the signal and its position in reciprocal space confirm that its source is an out-of-plane CDW.

Bragg scans along the (0,0,$L$) direction taken 400 μm into the capped region as shown in Figure 2(b) revealed no (0,0,2-2$\delta$) CDW peak at 4 K (red curve) or at RT (blue curve). The oscillations observed at both temperatures in the capped region were interference fringes arising from the thickness of our Fe layer. By using the relation between thickness and reciprocal space period, $t = 2\pi/\Delta q$, we determined that the Fe layer is 7.4 nm thick. The equivalence of the peaks at 4 K and RT is demonstrated by taking their difference (green curve). This indicates that there is no out-of-plane (0,0,2-2$\delta$) CDW signal in the capped region, and we conclude that ***Q*** lies parallel to the film plane in this region. Therefore we have created a 90º ***Q*** domain wall at the capped-uncapped boundary defined by our lithographic mask.

The region around the domain wall was characterized with x-ray microdiffraction. Using a sub-500 nm focused x-ray beam, a series of radial scans were taken along (0,0,*L*) at each of 18 one-micron-sized steps across the sample, extending from 2.5 μm into the capped side to 14.5 μm into the uncapped side. Figure 3(a) shows this data after subtracting the temperature independent background to reveal the CDW signal. The fluorescence signal of Fe was also collected at each point and is shown in Figure 3(b) along with the summed microdiffraction intensity from the CDW signal. A side-by-side comparison of the microdiffraction and microfluorescence data confirmed the presence of the (0,0,2-2$\delta$) CDW domain at (0,0,1.901) in the uncapped side and absence of this CDW satellite in the capped side. The CDW domain wall was determined to be less than 1 micron wide and occurs precisely at the mask-defined boundary between Fe-capped and uncapped sides.

In summary, we have for the first time engineered antiferromagnetic domain wall at a specific location in Cr thin film using lithographic patterning of ferromagnetic cap layers. The boundary between capped and uncapped layers serves as an artificial pinning center for domain wall nucleation within a Cr film. We have shown that an antiferromagnetic domain wall may be created and positioned at the demarcation line of our lithographically defined Fe mask, and that the antiferromagnetic domain wall width is narrower than our step size of 1 micron. By varying the shape, size, and location of patterned features, it is possible to produce antiferromagnetic domain walls precisely tailored to the technological demand. Furthermore, it is known that at low temperatures, the SDW of Cr films is longitudinally polarized, with ***S*** and ***Q*** pointing along the same direction [13,14]. This method, therefore, makes possible the engineering of both antiferromagnetic propagation direction and spin polarization domain walls at temperatures below 40 K. The technique introduced here will be useful in future investigations of domain wall interactions with pinning centers as well as transport measurements across individual antiferromagnetic domain walls. We foresee many applications of this method for technology as the role of antiferromagnets in spintronic and other magnetic devices continues develop and expand.

Films were grown, patterned and etched using facilities of the Center for Nanoscale Materials at Argonne National Laboratory. X-ray data were collected at beamline 2ID-D of the Advanced Photon Source, Argonne National Laboratory. Use of the Center for Nanoscale Materials and the Advanced Photon Source were supported by the U. S. Department of Energy, Office of Science, Office of Basic Energy Sciences, under Contract No. DE-AC02-06CH11357. O.S. would like to acknowledge support by the U.S. Department of Energy, Office of Science, Office of Basic Energy Sciences, under Contract DE-SC0001805.

1. B. G. Park, J. Wunderlich, X. Marti, V. Holy, Y. Kurosaki, M. Yamada, H. Yamamoto, A. Nishide, J. Hayakawa, H. Takahashi, A. B. Shick, and T. Jungwirth. Nature Mater. 10, 347 (2011).
2. R. K. Kummamuru, and Y. A. Soh, Nature 452, 63 (2008).
3. R. Jaramillo, T.F. Rosenbaum, E.D. Isaacs, O.G. Shpyrko, P.G. Evans, G. Aeppli, and Z. Cai, Phys. Rev. Lett. 98, 117206 (2007).
4. P. G. Evans, E. D. Isaacs, G. Aeppli, Z. Cai, and B. Lai, Science 295, 1042 (2002).

5. R. P. Michel, E. E. Israeloff, M. B. Weissman, J. A. Dura, and C. P. Flynn, Phys. Rev. B 44, 7413-7425 (1991).

6. O. G. Shpyrko, E. D. Isaacs, J. M. Logan, Y. Feng, G. Aeppli, R. Jaramillo, H. C. Kim, T. F. Rosenbaum, P. Zschack, M. Sprung, S. Narayanan, and A. R. Sandy, Nature, 447, 68 (2007).

7. E. Fawcett, Rev. Mod. Phys. 60, 209-283 (1988).

8. H. J. Zabel, Phys. Condens. Matter 11, 9303 (1999).

9. P. Sonntag, P. Bödeker, T. Thurston, and H. Zabel, Phys. Rev. B 52, 7363 (1995).

10. P. Bödeker, A. Hucht, A. Schreyer, J. Borchers, F. Güthoff, and H. Zabel, Phys. Rev. Lett. 81, 914 (1998).

11. E. E. Fullerton, M. J. Conover, J. E. Mattson, C. H. Sowers, and S. D. Bader, Appl. Phys. Lett. 63,1699 (1993).

12. J. P. Hill, G. Helgesen, and D. Gibbs, Phys. Rev. B **51** 10336 (1995).

13. P. Bodeker, A. Schreyer, and H. Zabel, Phys. Rev. B **59**, 9408 (1999).

14. P. Sonntag, P. Bodeker, A. Schreyer, H. Zabel, K. Hamacher, and H. Kaiser, J. Magn. Magn. Mater. 183, 5 (1998).

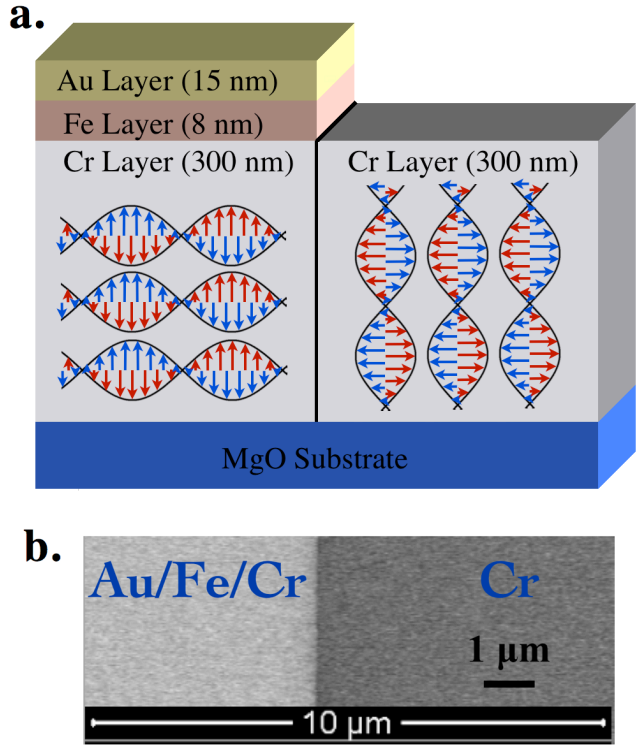

FIG. 1. (Color Online) (a.) Schematic (oblique side view) of the epitaxial trilayer film used in this experiment showing the orientation of the transverse spin-density wave (SDW) and accompanying charge-density wave (CDW). The *capped* Cr layer on the left shows only one of two possible orthogonal orientations for the in-plane CDW. In the *uncapped* region of Cr we expect a single out-of-plane orientation of the CDW following the removal of Fe. (b.) Plan-view SEM image of the boundary between the Fe *capped* and *uncapped* sides of the Cr film.

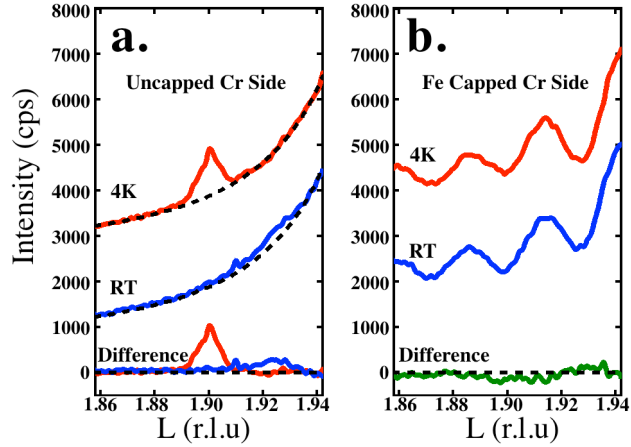

FIG. 2. (Color Online) X-ray scans along the (0,0,*L*) direction with an unfocused (200 x 200) μm² beam. (a.) Scans taken 400 microns into the *uncapped* Cr region. The 4 K data (top red curve) is displaced upward by 2000 cps for clarity. The top blue curve shows RT data. The background (dashed black line) has been fitted with the same Gaussian function for both curves. The bottom red and blue curves represent the 4 K and RT data, respectively, with the background subtracted. Background subtraction leaves only the out-of-plane (0,0,2-2$\delta$) CDW intensity centered at 1.901 at 4 K (red) and around 1.923 at RT (blue). (b.) Scans taken 400 microns into the capped Fe region. The peaks are interference fringes due to the thickness of the Fe cap layer. The 4 K data (red curve) is displaced upward by 2000 cps for clarity. The blue curve shows data taken at RT. The dark green curve is the difference between 4 K and RT showing that the signal is temperature independent. There is no signal arising from an out-of-plane (0,0,2-2$\delta$) charge-density wave on the Fe capped side.

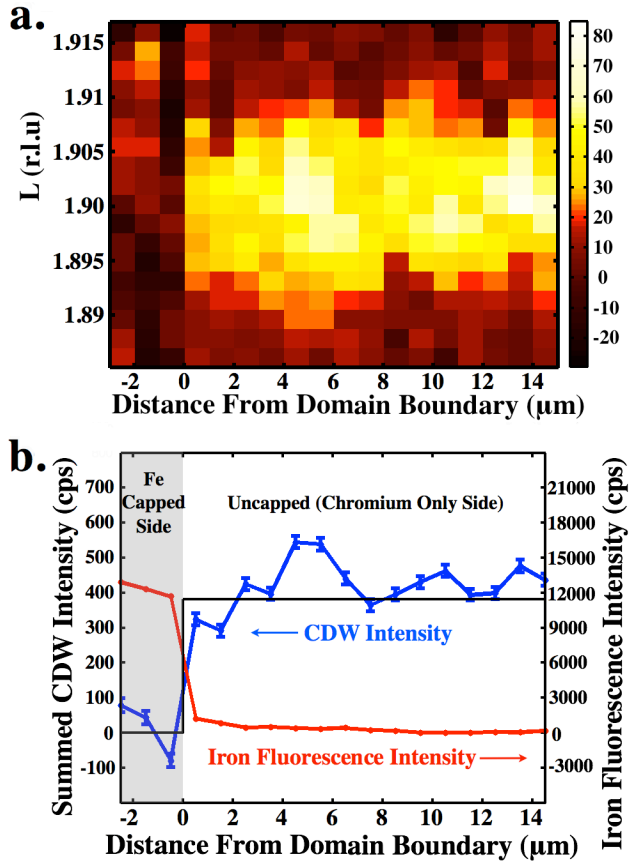

FIG. 3. (Color online) (a.) X-ray microdiffraction scan at 4 K across the Fe capped/uncapped boundary of the Cr film where positive (negative) distance corresponds to the distance from the domain wall into the uncapped (capped) side. Scan shows CDW intensity along the (0,0,$L$) direction after subtracting the temperature independent background. The increased signal around Q=1.901 r.l.u in the uncapped side is expected for the out-of-plane CDW of chromium at 4 K. In the capped region, scans show no peak indicating that the CDW has rotated to lie in-plane. (b.) Fe x-ray microfluorescence intensity (red curve) and summed CDW intensity (blue curve) across the capped/uncapped boundary. The black curve is the expected CDW intensity predicted from the unfocused beam data of Fig. 2.